\documentstyle[preprint,tighten,prl,aps,epsfig,epsf]{revtex}

\begin{document}


\title{Modification of the valence band electronic structure under\\
Ag intercalation underneath graphite monolayer on Ni(111)}

\author{Yu. S. Dedkov, M. Poygin, D. Vyalikh,\\ A. Starodubov, A. M. Shikin, and V. K. Adamchuk}
\address{Institute of Physics, St. Petersburg State University,\\
         198904 St. Petersburg, Russia}

\maketitle

\begin{abstract}
Angle-resolved photoemission spectroscopy and Auger electron
spectroscopy have been applied to study the intercalation process
of silver underneath a monolayer of graphite (MG) on Ni(111). The
room-temperature deposition of silver on top of MG/Ni(111) system
leads to the islands-like growth of Ag on top of the MG. Annealing
of the "as-deposited" system at temperature of 350-450$^\circ$\,C
results in the intercalation of about 1-2\,ML of Ag underneath MG
on Ni(111) independently of the thickness of pre-deposited Ag film
(3-100\,\AA). The intercalation of Ag is followed by a shift of
the graphite-derived valence band states towards energies which
are slightly larger than ones characteristic for pristine
graphite. This observation is understood in terms of a weakening
of chemical bonding between the MG and the substrate in the
MG/Ag/Ni(111) system with a small MG/Ni(111) covalent contribution
to this interaction.
\end{abstract}

\pagebreak {\bf Corresponding author:}

Yu. S. Dedkov

II. Physikalisches Institut

52056 Aachen, Germany

Tel.: ++49-241-8027091

Fax: ++49-241-8022306

E-mail: dedkov@physik.rwth-aachen.de

\pagebreak

\section{Introduction}
The continuing interest in carbon-based material has resulted in a
number of studies of their magnetic, electronic, and transport
properties which are interesting from technological as well as
fundamental point of view~\cite{Dresselhaus,Gorovikov}. The
ability to alter the properties of graphite and graphite
intercalation compounds (GICs) makes them interesting candidates
for a range of applications including, for example, magnetic
storage, hard coating, and optical devices. Also,
quasi-two-dimensional or even one-dimensional character of
mentioned structures gives the opportunity to use them as model
systems for different studies in solid-state physics and
chemistry.

In the last decades the systems on the basis of noble metals (Cu,
Au, Ag) intercalated underneath a monolayer of graphite (MG) on
top of Ni(111) were extensively investigated by means of
high-resolution-electron-energy-loss-spectroscopy
(HREELS)~\cite{Shikin:1998,Shikin:1999,Farias:1999,Shikin:2000}
and photoemission spectroscopy
(PES)~\cite{Shikin:2000,Dedkov:2001} methods. These systems have
attracted a strong interest due to a quasi two-dimensional
geometric and electronic structure. In these noble metal
intercalation-like systems an unusual effect of energetic shift of
graphite-derived branches in the valence band electronic structure
and the phonon vibration spectra after intercalation of noble
metals were observed. This effect is caused by the weakening of
the "MG-substrate" interaction due to the noble metal
intercalation. From the other side, this intercalation-like system
on the basis of MG and noble metal can be chosen as a possible way
of formation of a flat layered structure, in spite of the case of
noble metal deposition on pure Ni(111) surface, where the
formation of three-dimensional island-like structures was
observed. In particular, in work~\cite{Koschel:1999} the formation
of flat pseudomorphic copper layer on top of Ni(111) was possible
by using a benzene as surfacant with subsequent removing of
benzene from the surface after intercalation process.

In this paper the investigations of the valence band electronic
structure modification  upon Ag intercalation underneath MG
located on top of the Ni(111) by angle-resolved photoemission
spectroscopy (ARPES) are presented. This work is a continuation of
HREELS-investigations of this system~\cite{Farias:1999}. In spite
of our previous PES investigations of the analogous systems on the
basis of copper and gold as
intercalant~\cite{Shikin:2000,Dedkov:2001}, in the current work,
the attention was attracted to the changes in the graphite-derived
electronic structure as well as to estimation of amount of
intercalated Ag as a dependence of the annealing temperature as
well as of the amount of deposited Ag on top of MG/Ni(111) system.

The present investigations show that the most appropriate
temperature for intercalation of Ag atoms underneath MG on Ni(111)
is in the range between 350 and 400-450$^\circ$\,C. Above these
temperatures the partial system degradation  was observed. As a
result of Ag atoms intercalation at 350-450$^\circ$\,C the
energetic shift of the graphite-derived valence band states was
observed. In this case these branches are located at lower binding
energies in comparison with the MG/Ni(111) system that can be
explained by the weakening of the "MG-substrate" interaction after
Ag intercalation underneath MG.

As it was mentioned above one of the present work tasks was
determination of the intercalated amount of Ag underneath MG on
top of Ni(111). In the recent works~\cite{Tontegode1,Tontegode2}
the process of Ag atoms intercalation underneath MG on Ir(111)
textured foil was investigated. Auger electron spectroscopy (AES)
and thermo-desorption spectroscopy (TDS) investigations showed
that maximal amount of Ag which can be intercalated underneath MG
is one monolayer (ML) in contrast to the Ag concentration
deposited on top of MG. Earlier in our previous
work~\cite{Farias:1999} the assumption about intercalation of only
1\,ML of Ag underneath MG and that the rest of deposited Ag formed
a large 3D islands on top of MG/Ni(111) system was made.

To make this problem clear in the present work the AES experiments
for three different amount of pre-deposited Ag (3, 10 and
100\,\AA) were performed. For estimation of Ag intercalated amount
the mean free path of AES electrons should be taken into account.
In this case for careful analysis of the intercalation process the
high kinetic energy (E$_{kin}$) AES lines of elements were chosen.
As a result of AES investigations the amount of Ag intercalated
underneath MG was estimated to about 2\,ML independently on
thickness of pre-deposited amount of Ag (3-100\,\AA). This value
is very small in comparison to the amount of Ag deposited on the
sample surface before the intercalation. In this case the
formation of large 3D islands of non-intercalated Ag is expected.

\section{Experimental details}
The experiments were carried out in two separated UHV experimental
chambers equipped with low energy electron diffraction (LEED) and
AES optics, an ion gun and a gas inlet system. Angle-resolved
photoemission measurements were carried out with ADES-type
spectrometer at the Berliner Electronenspeicherring f\"ur
Synchrotronstrahlung (BESSY I, Berlin, Germany) using a
monochromatic light with the photon energy of 50\,eV from TGM2
beamline. Total energy resolution was 100\,meV and angle
resolution was 1$^\circ$. The AES experiments were carried out by
four-grid electron spectrometer with relative energy resolution of
0.25\%. All spectra were collected in $dN/dE$ mode with energy of
primary electron beam of 1000\,eV and modulation voltage of 2\,V.
The base pressure in the experimental chambers was better than
1$\times$10$^{-10}$\,mbar.

MG was formed on top of the Ni(111) surface by cracking of
propylene (C$_3$H$_6$) at surface temperature of 500$^\circ$C
using the recipe described in details
in~\cite{Shikin:1998,Shikin:1999,Farias:1999,Shikin:2000,Dedkov:2001}.
Sample exposition at a propylene pressure of
1$\times$10$^{-6}$\,mbar during 4 min was enough to form a well
ordered MG layer on top of Ni(111) surface. The crystallographic
structure was checked by LEED and ordered hexagonal LEED picture
with three-fold symmetry was
observed~\cite{Shikin:1998,Shikin:1999,Farias:1999,Shikin:2000,Dedkov:2001}.
Ag films were deposited at room temperature from melted Ag piece
(purity 99.9\%) supported by W wire which was heated by current
and thickness of the deposited layer was simultaneously measured
by water-cooled quartz microbalance monitor. The processes of Ag
intercalation underneath MG on top of Ni(111) surface was
investigated at a substrate temperature in the range between 350
and 500$^\circ$\,C. Same sample preparation procedure of the
intercalation-like system on the basis of Ag and MG was described
earlier in our work~\cite{Farias:1999} for HREELS investigations
of the phonon structure of MG/Ag/Ni(111) system.

\section{Experimental results and discussion}

In Fig.1 a series of the normal emission valence band
photoemission spectra is presented for "as-deposited"
100\,\AA-thick Ag-overlayer on top of the MG/Ni(111) system as
well as after annealing of the (Ag + MG/Ni(111)) system at
different temperatures in the range between 350 and
500$^\circ$\,C. These temperature regimes are appropriated for
Ag-intercalation in accordance with our previous HREELS
investigations~\cite{Farias:1999}. For comparison, the
corresponding spectrum for the initial MG/Ni(111) system is shown
by open symbols.

PES spectrum for the MG/Ni(111) system is characterized by
pronounced $\pi_{1v}$ and $\sigma_{2,3v}$ graphite-derived
features located at the binding energies (BE) of about 10.2 and
5\,eV, respectively, and two Ni 3\textit{d}-derived features with
maxima at 0.5 and 1.4\,eV BE. Room-temperature deposition of
100\,\AA\ Ag leads to suppression of MG- and Ni-derived states in
the PES spectra. The dominant features in the spectra are Ag
4\textit{d}-derived states located in the BE range between 4 and
7\,eV. Despite a significant amount of deposited Ag (about of
100\,\AA) the weaked Ni 3\textit{d} states at the BE of 0.5 and
1.5\,eV and very weak $\pi_{1v}$ graphite-derived features at
about of 10\,eV can be distinguished in the spectra that can
testify an island-like character of "as-deposited" Ag-layer.
Additional feature observed at the Fermi-level (E$_{F}$) can be
related to the surface states characterized for metallic
silver~\cite{Kevan:1987}.

Annealing of the 100\,\AA\,Ag/MG/Ni(111) system at 350$^\circ$\,C
is followed by the restoring of the $\pi_{1v}$ graphite-derived
states and disappearing of the Ni 3\textit{d} states in the energy
region between E$_F$ and 1.5\,eV BE. In comparison with the
initial MG/Ni(111) system the $\pi_{1v}$ graphite-derived states
are shifted towards lower BE by of 1.3\,eV up to of about 9\,eV
BE. The Ag 4\textit{d}-derived features remain dominant ones in
the PES spectra. Increase of annealing temperature up to 400 and
450$^\circ$\,C leads to further decreasing of the BE of
$\pi_{1v}$-derived states. After annealing at 450$^\circ$\,C the
BE of the $\pi_{1v}$ graphite-derived states reaches the value of
about 8.3\,eV relative to E$_F$. Further increase of annealing
temperature up to 500$^\circ$\,C is followed by reverse increasing
of  the BE of the $\pi_{1v}$-derived states. The structure of the
Ag 4\textit{d}-derived states does not changed at these annealing
temperatures. The restoring of the intensity of $\pi_{1v}$
graphite-derived states and the decreasing of their BE after
annealing of the Ag/MG/Ni(111) system at temperatures above than
350$^\circ$\,C can be explained by the intercalation of Ag atoms
underneath MG on top of Ni(111) and the change of the
"MG-substrate" interaction after intercalation.

In Fig.2 the changes of the relative intensities of the $\pi_{1v}$
graphite-derived states at 10.2\,eV BE and the Ag
4\textit{d}-derived feature located at 4.5\,eV BE are presented
for the annealed system at different temperatures in comparison
with those for the MG/Ni(111) system and for the "as-deposited" Ag
overlayer on top of MG/Ni(111) system. It can be seen that
annealing at 350$^\circ$\,C leads to a growth of the intensity of
the $\pi_{1v}$ graphite-derived feature and suppression of the
intensity of the Ag 4\textit{d}-derived one. Increase of the
annealing temperature up to 400 and 450$^\circ$\,C leads to the
further increasing of the intensity of the $\pi_{1v}$
graphite-derived states and suppression of the Ag
4\textit{d}-derived states. However, annealing at temperature of
500$^\circ$\,C is followed by the decreasing of intensity of the
$\pi_{1v}$ graphite-derived states as well as by the increasing of
the Ag 4\textit{d}-derived ones.

On the basis of the presented data (changes of the BEs and of the
intensity of the graphite- and Ag 4\textit{d}-derived states) the
optimal temperature for the  Ag-intercalation in the range between
350 and 450$^\circ$\,C was found. At higher annealing temperature
leads to the start of the MG/Ag/Ni(111) system degradation process
was observed.

In order to estimate the quantity of intercalated Ag underneath MG
on top of the Ni(111) surface the AES experiments were carried
out. Three different thicknesses of Ag were used (3, 10 and
100\,\AA). The AES intensities of the C(KVV),
Ag(M$_5$N$_{45}$N$_{45}$), Ni(L$_3$M$_{45}$M$_{45}$) and Ni (MVV)
lines with E$_{kin}$=272, 351, 848, and 60\,eV respectively,
depending on annealing temperature for the case of 100\,\AA\ of
pre-deposited Ag on top of MG/Ni(111) system are shown in Fig.3a.
The choice of these lines for quantitative analysis was caused by
the possibility to carefully separate the contributions from
different elements (C, Ag and Ni) and by the reason that the
electrons with such kinetic energy have a larger mean free path
than electrons with E$_{kin}$ around 50\,eV analysed in PES
experiments. Encircled numbers I and II in Fig.3a correspond to
MG/Ni(111) system and system after deposition of corresponding
amount of Ag on top of MG, respectively. After Ag deposition the
intensities of the C and Ni AES peaks decreased and Ag AES peak
increased. After annealing of the Ag/MG/Ni(111) system the
situation is changed: intensity of the Ag AES peak is decreased
and intensities of the carbon and Ni peaks are increased. At
annealing temperature of about 300$^\circ$\,C the intensity of
carbon peak is the same as for the MG/Ni(111) system. In this case
the situation is following: the graphite layer is on the top of
the system and Ag intercalation underneath MG on top of Ni(111)
takes place. In the same time the Ni AES peak has a visible
intensity at least three times smaller in comparison with the
MG/Ni(111) system.

For comparison, the changes of the intensity ratio of the Ni and
Ag AES peaks with temperature change are presented in Fig.3b and
marked by solid squares, circles and triangles for 3, 10 and
100\,\AA\ of pre-deposited Ag, respectively.  For systems with 3
and 10\,\AA\ of pre-deposited Ag on top of the MG/Ni(111) the
behaviours of the AES spectra with changes of annealing
temperature are the same (not shown here): decreasing of the
intensity of the carbon and Ni AES peaks after deposition of Ag
and increasing ones after annealing at temperatures above
250$^\circ$\,C. For all three cases the intensity of the carbon
AES peak for the system, which is produced by the annealing of the
Ag/MG/Ni(111) system at 350-370$^\circ$\,C, is nearly the same and
has an intensity comparable with intensity of carbon AES peak for
the MG/Ni(111) system. In all cases the Ni AES peaks were also
observed. From the analysis of the Fig.3a and 3b the main features
can be pointed: (i) intensity of the Ni AES peak for the case of
100\,\AA\ pre-deposited Ag after annealing procedure is two times
smaller than for the case of 3\,\AA\ pre-deposited Ag, where
intercalation of 1\,ML of Ag was supposed, (ii) for the system
with 100\,\AA\ of pre-deposited Ag the intensity of the Ni AES
peak after annealing is decreased by three times in comparison
with MG/Ni(111) system, (iii) for all three systems the Ni AES
peak with kinetic energy of 60\,eV is clearly visible (this peak
has a small inelastic mean free path in comparison with other AES
peaks). On the basis of this analysis one can conclude that for
all three cases of pre-deposited Ag on top of MG/Ni(111) system,
only about 2\,ML of Ag can be intercalated underneath MG on top of
Ni(111) independently on thickness of pre-deposited Ag film. This
conclusion is in a good agreement with results of
works~\cite{Farias:1999,Tontegode1,Tontegode2} where the thickness
of intercalated Ag was estimated as 1\,ML, but near some
imperfections of the surface of the Ni crystal (as a defects,
steps, dislocations etc.) the intercalations of two or more
monolayers of Ag can be expected.

For analysis of the peaks intensities extracted from PES data the
mean free path of photoelectrons at $h\nu$=50\,eV has to be taken
into account. From these estimations of the intensity changes of
Ni 3\textit{d} states near E$_F$ one can conclude that 2-3\,MLs of
covered layer on top of Ni(111) (together with MG) should be
enough to completely depress the intensity of these states near
E$_F$. These estimations together with presented AES data lead to
a conclusion that in case of Ag intercalation underneath MG only
1-2 monolayers of Ag are intercalated underneath MG matrix. The
rest of Ag can decorate the defects at the surface and form a
randomly-situated large 3D-islands  as a lot of dislocations were
observed by STM on MG surface as it was shown in previous
investigations of copper intercalations underneath
MG~\cite{Dedkov:2001}. The next possibility is the agglomeration
of Ag at the borders of the sample. This estimation can be
compared with the results of work~\cite{Shikin:2000,Dedkov:2001}.
In case of Au the 1\,ML intercalation underneath MG on top of
Ni(111) the Ni 3\textit{d} states near E$_F$ were also fully
depressed. In opposite, for Cu intercalation the Ni 3\textit{d}
states still can be distinguished. Such diferences in the Ni
3\textit{d} states behaviour for the noble metal intercalation
underneath MG on Ni(111) can be explained by the nickel-copper
alloy formation and absence of such alloys for Au and Ag at
temperatures essential for noble metal intercalation.

In Fig.4 a series of angle-resolved PES spectra for the
MG/Ag/Ni(111) system formed by annealing of 100\AA\,Ag+MG/Ni(111)
system at 450$^\circ$\,C are presented. Spectra were measured in
$\overline{\Gamma}$-$\overline{K}$ direction of the graphite
surface Brillouin zone (SBZ). Corresponding dispersions of the
main valence band graphite-derived electronic states are shown in
Fig.5 for the systems formed by annealing at temperatures of 350,
400 and 450$^\circ$\,C. Also in Fig.5 the dispersion in
$\overline{\Gamma}$-$\overline{M}$ direction for the MG/Ag/Ni(111)
system with the 3\,\AA\ and 4.7\,\AA\ of the pre-deposited Ag is
shown by open squares and circles, respectively. The measurements
in this case were made with the He resonance lamp and photon
energy was 40.8\,eV. Here, for comparison the dispersion of the
$\pi$ and $\sigma$ graphite-derived states for the MG/Ni(111)
system as well as for pristine graphite are shown by straight and
dash-dotted lines, respectively. Similar to Fig.1 the PES spectra
presented in Fig.4 are characterized by dominant contribution of
the Ag 4\textit{d}-derived features. The fine structure of the Ag
4\textit{d}-derived states practically do not changed with the
polar angle variation. These states are practically non-dispersive
in the PES spectra. Besides the Ag 4\textit{d}-derived structure
the PES spectra are characterized by the $\pi_{1v}$, $\sigma_{1v}$
and $\sigma_{2,3v}$ graphite-derived states with strong
dispersion. Binding energy of the $\sigma_{1v}$ and $\pi_{1v}$
states in the $\overline{\Gamma}$ point of the SBZ (normal
emission) are equal to 21.5 and 8.3\,eV, respectively. By increase
of the polar angle the $\pi_{1v}$-states reach the position with
the BEs lower than those for the Ag 4\textit{d}-derived states.
For polar angles over 25$^\circ$ ($k_{||}>$1.4-1.5\,\AA$^{-1}$)
the $\pi$-states are splited in two features. One feature reaches
the BE of about 3\,eV for $k_{||}\sim$1.5\,\AA$^{-1}$. Second
feature has the lowest BE equals of about 1\,eV for the polar
angles in the region of 30$^\circ$ ($k_{||}=$1.7\,\AA$^{-1}$) that
corresponds to the $\overline{K}$ point of graphite SBZ. By
further increase of polar angle both states show decreasing
intensities and shift towards higher BEs. Besides the $\pi_{1v}$
and $\sigma_{1v}$ graphite-derived states, the $\sigma_{2,3v}$
graphite-derived states can be distinguished in the PES spectra
for polar angles over 20$^\circ$(Fig.5). These states are also
strongly dispersive. Dispersive dependencies for these states are
presented in Fig.5. In comparison with the dispersion dependencies
of the MG/Ni(111) system the branch of the $\pi_{1v}$
graphite-derived states is significantly shifted as whole towards
lower BEs by about of 1.5-2\,eV after intercalation of Ag atoms
underneath MG on Ni(111). In the $\overline{K}$ point of the
graphite SBZ the BE of the $\pi_{1v}$ states for the MG/Ag/Ni(111)
system is about of 1\,eV in comparison with the BE about of 3\,eV
for the MG/Ni(111) system. Totally  the $\pi_{1v}$ states for the
MG/Ag/Ni(111) system are located in the region between branches
characteristic for pristine graphite and for the MG/Ni(111)
system. In the region of the $\overline{K}$ point they are located
more close to the positions characteristic for pristine graphite.
The $\sigma_{1v}$ states in the region of the $\overline{\Gamma}$
point are located at the energy characteristic for pristine
graphite. In the region of the $\overline{K}$ point these states
have the BE about of 16.5\,eV. This energetic shift of the
graphite-derived valence band states for the MG/Ni(111) system can
be attributed to weakening of the interaction of MG with
Ni-substrate after intercalation of Ag atoms. In accordance
to~\cite{Nagashima:1994,Oshima:1997} this interaction for the
MG/Ni(111) system is caused by orbital mixing (hybridization) of
the $\pi$-states of MG with 3\textit{d}-states of the substrate.
Such hybridization is followed by electron redistribution from the
bonding $\pi$-states to antibonding $\pi^{*}$-states that results
in the weakening of the (C-C) bonds in the graphite layer and
increase of the BE of the graphite-derived states (different for
the $\sigma$ and $\pi$ states). Intercalation of Ag atoms
increases the distance between MG and Ni-substrate and, as
consequence, leads to decreasing of the MG $\pi$ - Ni 3\textit{d}
interaction and to significantly lower BEs of the $\pi$ and
$\sigma$ graphite-derived states as compared to those in the
MG/Ni(111) system. The assumption about the weakening of the
"MG-substrate" interaction after Ag-intercalation is in agreement
with the results of HREELS investigations of the MG/Ag/Ni(111)
system~\cite{Farias:1999} which displays the analogous effect of
"stiffening" of the graphite-derived phonon modes after
Ag-intercalation.

In some degree analogous energetic shift of the graphite-derived
states were recently observed upon intercalation of gold and
copper underneath MG on the
Ni(111)~\cite{Shikin:2000,Dedkov:2001}. But in that case the $\pi$
and $\sigma$ graphite-derived states were shifted upon Cu or Au
intercalation up to the positions characteristic for pristine
graphite. Such difference between modification of the PE spectra
upon intercalation of Ag, Cu and Au can be explained by more
complex interaction of Ag with MG and Ni-substrate. According to
results of Oshima \textit{et al.} calculations of the phonon modes
in the MG/Cu/Ni and MG/Ag/Ni systems the interaction of MG with
substrate after intercalation of Ag is more stronger than that,
which realizes in the case of pristine graphite and the
Mg/Cu/Ni(111) system~\cite{Farias:2000}.

On the basis of these experimental observations the following
model can be proposed: after intercalation of Ag in the MG/Ni(111)
system the covalent-like nature of interaction between MG and
substrate is saving due to the (Ni - Ag) interaction. However, the
strength of this interaction is lower than in the case of the
MG/Ni(111) system. This assumption is supported by the
experimental fact, which shows that the occupation of the
$\pi^{*}$ graphite-derived states and the appearance corresponding
features in the PES spectra were not observed. This situation is
opposite to the case of the alkali metal intercalation underneath
MG where these effects were observed~\cite{Nagashima:1994}.

Other interesting effect observed after the Ag-intercalation is
the splitting of the $\pi$ graphite-derived states in the PES
spectra for polar angles over 25$^\circ$
($k_{||}>$1.4\,\AA$^{-1}$). In our opinion the observed difference
in the energetic positions and the angle dependencies of both
$\pi$ grahite-derived branches maxima can be related to different
behavior of the $\pi$ states in the regions of the $\overline{K}$
and $\overline{M}$-point of the graphite SBZ. On this basis one
can assume that the second branch of the $\pi$ states with maximum
of the BE at 4\,eV at $k_{||}\sim$1.4\,\AA$^{-1}$ is similar to
the behavior of the $\pi$ states in the region of the
$\overline{M}$-point of the SBZ, i.e. it can be the "footprint" of
the $\overline{\Gamma}-\overline{M}$ direction (see, for
comparison, the real dispersion in
$\overline{\Gamma}-\overline{M}$ direction shown by open symbols).
It can be caused by some domains disorientation in graphite
monolayer which can arrive after Ag atoms intercalation. The BE
value of the $\sigma_{1v}$-states observed in the region of the
$\overline{K}$-point of the SBZ can be also understood from
analogous positions. This part of the $\sigma_{1v}$
graphite-derived branch can be also related to "footprint" of the
$\sigma_{1v}$ states in the $\overline{\Gamma}-\overline{M}$
direction. The LEED pattern of the system after Ag atoms
intercalation displays graphite-like (1$\times$1) hexagonal
structure characteristic for the MG/Ni(111) system with some weak
strips rotated on 15$^\circ$ with respect to hexagonal ones. Such
LEED patterns suggest that the prepared system is terminated by
graphite-like domains with some orientation misfit with respect to
the Ni(111) substrate, as it has been already observed for the
MG/Cu/Ni(111) system~\cite{Shikin:1998,Shikin:1999} and has been
described for the MG/Ag/Ni(111) system~\cite{Farias:1999}. This
disorientation is ordered one that leads to equivalent electronic
structure both in the $\overline{\Gamma}-\overline{K}$ and
$\overline{\Gamma}-\overline{M}$ directions of the monocrystalline
graphite SBZ. From the other side the splitting of the $\pi$
graphite-derived states for the MG/Ni(111) system around
$\overline{K}$-point of the graphite SBZ was theoretically
predicted by Souzu \textit{et al}.~\cite{Souzu:1995}. The energy
dispersion  of the $\pi$ graphite-derived states along the
high-symmetry line $\overline{\Gamma}-\overline{K}-\overline{M}$
has a splitting-point at around of 2\,eV BE. But for the isolated
graphene layer which can be taken as an ideal case in the model of
MG/Ag/Ni(111) intercalation-like system this splitting-point lies
at E$_F$~\cite{Souzu:1995}. The possible energy shift can be
caused by the rest interaction MG with substrate and deviation
from the ideal model of isolated graphene layer.

\section{Conclusion}

On the basis of the PES and AES investigations of the Ag
intercalation process underneath MG on top of the Ni(111) surface
the following conclusions can be made: (i) optimal temperature for
intercalation of Ag underneath MG lies in the range
350-450$^\circ$\,C; at temperature above 500$^\circ$\,C the
degradation of the MG/Ag/Ni(111) system starts; (ii) the valence
band electronic structure of the MG/Ag/Ni(111) system is
characterized by the energetic shift of the $\pi$ and $\sigma$
graphite-derived states as compared with the MG/Ni(111) system
towards lower BEs by about of 1.5-2\,eV and 0.5-1\,eV,
respectively; it is assumed that the observed changes of
electronic structure is caused by the weakening of the
$\pi$-\textit{d} interaction of the MG with substrate; the
"orbital mixing" nature of this interaction is saving after Ag
intercalation; (iii) on the basis of intensity changes analysis of
C, Ag and Ni AES peaks in intercalation process a conclusion about
1-2\,ML of intercalated Ag underneath MG on top of Ni(111) was
made and this value practically independent from the amount of Ag
deposited on top of the MG/Ni(111) system before intercalation;
after intercalation of 1-2\,ML of Ag the formation of the large Ag
islands which have a total square less than 50\% of the surface
can be expected; the agglomeration of the Ag at the surface
dislocations and border of the sample can not be excluded.

\section{Acknowledgements}
This work was performed in framework of INTAS Project
No.\,2001-2136 and Program of Ministry of Industry and Science of
Russian Federation "Fullerenes and Atomic Clusters".




\pagebreak
  {\bf Fig.1.} PES spectra of the MG/Ni(111), 100\,\AA\ of Ag on MG/Ni(111), as well of Ag/MG/Ni(111) system
  after annealing at 350, 400, 450 and 500$^\circ$\,C, respectively.

  {\bf Fig.2.} The changes of relative intensities of the $\pi_{1v}$
  graphite-derived state and the Ag 4\textit{d}-derived feature
  after annealing of the Ag/MG/Ni(111) system at different temperatures in
  comparison with those for the MG/Ni(111) system and for the
  "as-deposited" Ag overlayer on top of MG/Ni(111) system.

  {\bf Fig.3.} a) Intensity of the C(KVV), Ag(M$_5$N$_{45}$N$_{45}$),
  Ni(L$_3$M$_{45}$M$_{45}$) and Ni(MVV) AES peaks in dependence from
  the annealing temperature for 100\,\AA\ of pre-deposited Ag on top of MG/Ni(111) system.
  b) temperature dependence of the intensity ratio
  (I(Ni$_{848}$)/I(Ag$_{351}$)) for different amount of pre-deposited Ag on
  top of the MG/Ni(111) system.

  {\bf Fig.4.} Series of the angle-resolved PES spectra measured at photon energy of 50\,eV for
  the MG/Ag/Ni(111) system formed after annealing of 100\AA\,Ag+MG/Ni(111) system at 450$^\circ$\,C.

  {\bf Fig.5.}Dispersion plot of the main valence band graphite
  derived electronic states for the Ag/MG/Ni(111) system formed
  after annealing of 100\,\AA\,Ag+MG/Ni(111) system at temperatures
  of 350 (up triangles), 400 (down triangles) and 450$^\circ$\,C
  (diamonds). The experimental data points for the
  $\overline{\Gamma}-\overline{M}$ direction and photon energy
  40.8\,eV are shown by open symbols. Here, for comparison the
  dispersion of the $\pi$ and $\sigma$ states for the MG/Ni(111)
  system as well as for pristine graphite are shown by straight and
  dash-dotted lines, respectively.

\end{document}